\documentclass[a4paper,12pt,dvips]{article}
\usepackage{amsmath}
\usepackage{amssymb}
\usepackage{graphicx}
\usepackage[dvips]{color}

\newcommand{\ii}{\mathrm{i}} 
\newcommand{\ee}{\mathrm{e}} 

\def\w{\!\wedge\!} 

\newcommand{\e}[1]{\mathbf{e}_{#1}} 
\newcommand{\Ie}[1]{I\!\mathbf{e}_{#1}}
\newcommand{\inv}[1]{\overline{#1}} 

\newcommand{\ba}{\mathbf{a}}
\newcommand{\bk}{\mathbf{k}}

\begin{document}

\title{Pseudospin, velocity and Berry phase in a bilayer graphene}
\author{A.~Dargys$^{*}$ and A.~Acus$^{**}$\\
\normalsize $^{*}$
Semiconductor Physics Institute, CPST,\\
\normalsize  A.~Go\v{s}tauto 11, LT-01108 Vilnius, Lithuania, dargys@pfi.lt\\
\normalsize  $^{**}$Institute of Theoretical Physics and
Astronomy, Vilnius University,\\
\normalsize  A.~Go\v{s}tauto 12, LT-01108 Vilnius, Lithuania,
acus@itpa.lt}

\normalsize\date{{\normalsize\today}}

\maketitle

\begin{abstract}
Hamiltonian and eigenstate problem is formulated for a bilayer
graphene in terms of Clifford's geometric algebra
\textit{Cl}$_{3,1}$ and respective multivectors. It is shown that
such approach allows to perform analytical calculations in a
simple way if geometrical algebra rotors are used. The measured
quantities are express through spectrum and rotation half-angle of
the pseudospin that appears in geometric algebra rotors.
Properties of free charge carriers -- pseudospin, velocity and
Berry phase -- in a bilayer graphene are investigated in the
presence of the external voltage applied between the two layers.
\end{abstract}


\section{Introduction}\label{intro}
Bilayer graphene (BLG) consists of two graphene monolayers A and B
that are typically  aligned in the Bernal  stacking arrangement.
The study of BLG has started  in 2006 after the appearance of
E.~McCann's and V.~I. Fal'ko's paper~\cite{McCann06}. A unique
feature of the BLG is its tunable energy band structure by
external voltage applied between A and B layers. The opening of
the gap in exfoliated bottom-gated BLG was demonstrated
experimentally by A.~B. Kuzmenko et al~\cite{Kuzmenko09}. A
comprehensive description of BLG properties and respective
references can be found  in review
articles~\cite{McCann13,Abergel10} and
monograph~\cite{Katsnelson12}.

The geometric algebra (GA), or more precisely a family of algebras
\textit{Cl}$_{p,q}$, where $p$ and $q$ define the metric of space,
can be used as an alternative mathematical instrument in
classical, quantum and relativistic
physics~\cite{Lounesto97,Doran03,Dorst11}. The main advantage of
GA is that it is coordinate or representation free, and unifies
all parts of applied mathematics (vector calculus, matrix algebra,
dyadics, tensor algebra, differential forms etc) in a single
coherent entity. Classical and quantum problems can be formulated
in the same space without any need to resort to abstract Hilbert
space. The GA has a very powerful tool called the ``rotor'' which
is related with quantum mechanical spinors and, therefore, with
the Schr{\"o}dinger and Dirac equations in multidimensional spaces
with variable metric~\cite{Hestenes92}. Recently, the GA and
rotors were applied to analyze quantum ring~\cite{Dargys13},
monolayer graphene and quantum
well~\cite{Bohmer12a,Dargys13a,Dargys14,Dargys14b}. Since
calculations in  GA are performed in a coordinate free way, the GA
is structurally compact and allows geometric interpretation of
formulas. This is very appealing as compared to abstract Hilbert
space approach. In this paper we apply relativistic
\textit{Cl}$_{3,1}$ algebra and its rotors to analyze bilayer
graphene properties. In the next section we calculate the rotors
and in subsequent sections the rotors are used to analyze the
pseudospin (Sec.~\ref{sec3}), velocity (Sec.~\ref{sec4}) and Berry
phase (Sec.~\ref{sec5}). In the Appendix, a summary of GA
operations and formulae used in this paper is presented.

\section{BLG in geometric algebra formulation}\label{sec2}
The BLG consists of two stacked carbon monolayers that  will be
numbered $1$ and $2$. The elementary cell of each monolayer has
two inequivalent sites $A$ and $B$, therefore, the wave function
in the tight binding approximation at least has four components
which will be arranged in the order
$\{\psi_{A1},\psi_{A2},\psi_{B1},\psi_{B2}\}$.  In this Hilbert
space basis  the considered BLG Hamiltonian is
\begin{equation}\label{hamBG}
\hat{H}=\eta\left[%
\begin{array}{cccc}
 U &0 & \gamma_0 k_{-} & 0\\
0& - U&\eta\gamma_1& \gamma_0 k_{-}\\
\gamma_0 k_{+} &\eta\gamma_1& U&0\\
0 &\gamma_0 k_{+}&0& -U
\end{array}%
\right]
\end{equation}
where $k_{\pm}=k_x\pm\ii k_y$ with $k_x$ and $k_y$ being the
components of wave vector  in the graphene plane. The constant
$\gamma_0=-\langle\psi_{A1}|\hat{H}|\psi_{B1}\rangle=-\langle\psi_{A2}|\hat{H}|\psi_{B2}\rangle$
is the intralayer  hopping energy and
$\gamma_1=\langle\psi_{A2}|\hat{H}|\psi_{B1}\rangle$ is the
interlayer nearest neighbor hopping energy. The energy  $2U$ is
the potential difference between the layers, also called the
asymmetry term. $\eta=+1$ for $K$-valley and $\eta=-1$ for
$K^{\prime}$-valley. The intrinsic electron spin in the
Hamiltonian~\eqref{hamBG}  is not included. In the following all
energies will be normalized to intralayer hopping energy
$\gamma_0$. The wave vector and its components will be normalized
by $k_c=2\pi/a$, where $a$ is the graphene lattice constant. It is
assumed that $\hbar=1$.

\begin{figure}
\centering
\includegraphics[width=5cm]{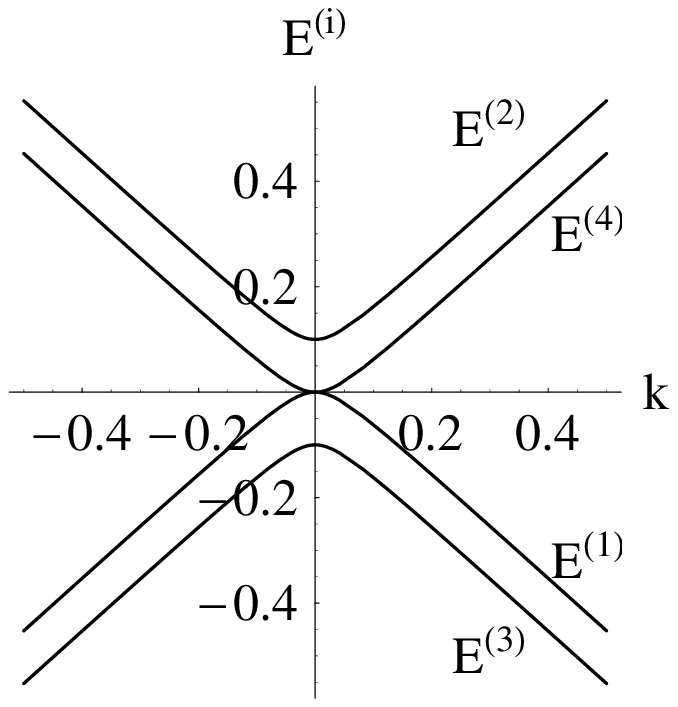}a)
\includegraphics[width=5cm]{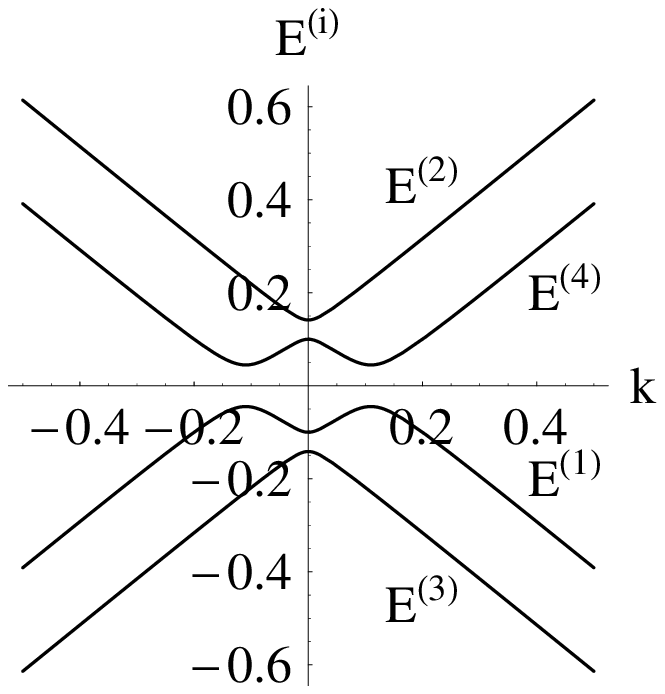}b)
\caption{The spectrum  of bilayer graphene, Eq.~\eqref{spectrum},
a) in absence, $U=0$ and b) in presence, $U=0.1$, of the
interlayer voltage $2U$. $\gamma_1=0.1$.}\label{f:spectra1234}
\end{figure}

The dependence of normalized to $\gamma_0$ dispersion energies of
the Hamiltonian~\eqref{hamBG} on the  wave vector magnitude
$k=(k_x^2+k_y^2)^{1/2}$ is
\begin{equation}\label{spectrum}
E^{(i)}=\pm\sqrt{k^2+U^2+\frac{\gamma_1^2}{2}\pm\frac{1}{2}\sqrt{\gamma_1^4+4k^2(4U^2+\gamma_1^2)}}\,,
\end{equation}
where $i$ is the band number as indicated in
Fig.~\ref{f:spectra1234}. The spectrum \eqref{spectrum}  is the
same for both $K$ and $K^{\prime}$ valleys. When $U=0$ the energy
gap is absent, Fig.~\ref{f:spectra1234}a, and the spectrum reduces
to $E_0^{(i)}=\pm(\gamma_1/2)\pm\sqrt{k^2+(\gamma_1/2)^2}$. If the
potential difference $2U$ is created between the layers the energy
gap opens  and parabolic-like spectrum of two  bands transforms to
``Mexican-hat''  spectrum as shown in Fig.~\ref{f:spectra1234}b.
This  brings about altogether new physical properties in BLG. Here
and in all subsequent figures the interlayer coupling energy is
assumed to be $\gamma_1=0.1$ which is close to experimental
values: $\gamma_1/\gamma_0=0.3\text{[eV]}/2.9\text{[eV]}\approx
0.1$ and $\gamma_1/\gamma_0=0.4\text{[eV]}/3.0\text{[eV]}\approx
0.13$~\cite{McCann13}.

Using matrix representation of \textit{Cl}$_{3,1}$ algebra  the
Hamiltonian~\eqref{hamBG} can be decomposed into sum of matrices
\begin{equation}\label{hatH}
\hat{H}=\eta(k_x\hat{e}_1+k_y\hat{e}_2)-\frac{\gamma_1}{2}\Big(\ii\,\hat{e}_{3}\hat{e}_{1}+
\hat{I}\hat{e}_2\Big)+\eta U\hat{I}\hat{e}_3,
\end{equation}
where  for $4\times 4$ matrices denoted by hats are given in the
Appendix~\ref{appendMat}. It should be stressed that the matrix
representation of GA elements is not unique and, it is important,
in principle is not needed in solving the problem at all. It is
presented for a convenience  to help the reader to connect the
present geometric algebra approach [see Eq.~\eqref{HBG} below]
with the standard Hilbert space Hamiltonian~\eqref{hamBG}.

\begin{figure}
\centering
\includegraphics[width=7cm]{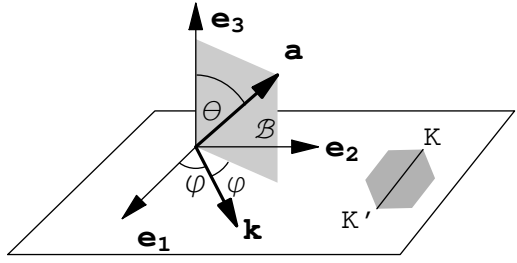}
\caption{Basis vectors $\e{1}$, $\e{2}$  and $\e{3}$, the wave
vector $\bk$, and the rotation plane
$\hat{\mathcal{B}}=\e{3}\w\hat{\ba}$ that contains $\e{3}$ and
vector $\ba$. The vector $\e{3}$ is perpendicular to graphene
plane (square). The line that connects $K$ and $K^{\prime}$
valleys in the Brillouin zone is parallel to
$\e{1}$.}\label{f:grapheneAxesab}
\end{figure}

Using the replacement rules from the Appendix~\ref{appendRules},
the matrix BLG Hamiltonian~\eqref{hatH} can be mapped to GA
Hamiltonian function of spinor $\psi$,
\begin{equation}\label{HBG}
H(\psi)=\eta\bk\psi\Ie{3}-\frac{\gamma_1}{2}
\e{2}\Big(\psi+\e{4}\psi\e{4}\Big)\e{3}+\eta U\e{3}\psi\e{3},
\end{equation}
where $\bk=k_x\e{1}+k_y\e{2}$ is the in-plane wave vector in terms
of basis vectors $\e{i}$ of \textit{Cl}$_{3,1}$ algebra. The
spinor $\psi$ that appears in this Hamiltonian function consists
of even grade blades only (see Eq.~\eqref{psipsi} in the
Appendix~\ref{appendRules}). In the Clifford space
Hamiltonian~\eqref{HBG}, the basis coordinate vector $\e{3}$ is
directed along pseudospin quantization axis,
Fig.~\ref{f:grapheneAxesab}.

Now  the eigenmultivector equation in GA
\begin{equation}\label{eigenEq}
H(\psi^{(i)})=E^{(i)}\psi^{(i)}
\end{equation}
is solved using the Hamiltonian~\eqref{HBG},  where $E^{(i)}$ is
the eigenenergy ($i$-th energy band) and $\psi^{(i)}$ is the
respective eigenmultivector, by the method proposed
in~\cite{Dargys13a,Dargys14}. The method is based on the fact that
the eigenvalue equation~\eqref{eigenEq} can be transformed to
rotor equation of the form $R\e{3}R^{-1}=\ba$, where the rotor $R$
brings the unit vector $\e{3}$parallel to the quasispin
quantization axis to a final position $\ba$ determined by the
Hamiltonian~\eqref{HBG}. The spectrum follows from the condition
that the rotation preserves the length of a rotated vector. The
respective eigenmultivectors then can be constructed from
bivectors that represent rotation planes which are made up of the
quantization axis and final direction determined by the
eigenvalue, Fig.~\ref{f:grapheneAxesab}.

To transform  the  Eq.~\eqref{eigenEq} to rotors, at first, the
multivector is divided into two parts $\psi=\psi_{+}+\psi_{-}$,
where $\psi_{+}$ is invariant to spatial inversion,
$\overline{\psi}_{+}=\psi_{+}$, and $\psi_{-}$ changes its sign to
opposite, $\overline{\psi}_{-}=-\psi_{-}$ (see the
Appendix~\ref{appendOp}). Then the multivector eigenequation
$H(\psi)=E\psi$, where the band index was temporally suppressed,
can be decomposed into pair of coupled multivector equations,
\begin{equation}\label{coupled}
\begin{split}
E\psi_{+}&=\eta\bk\psi_{-}\Ie{3}+\eta U\e{3}\psi_{+}\e{3},\\
E\psi_{-}&=\eta\bk\psi_{+}\Ie{3}+\eta
U\e{3}\psi_{-}\e{3}-\gamma_1\e{2}\psi_{-}\e{3}.
\end{split}
\end{equation}
The method of solution of~\eqref{coupled} relies on the fact that
$\psi_{+}$ and $I\psi_{-}$, where $I$ is the pseudoscalar of
\textit{Cl}$_{3,1}$ algebra, are the rotors in 3D Euclidean space.
For this purpose one expresses $\psi_{-}$ from the first equation
of~\eqref{coupled}, and inserts into the second equation. After
some algebraic manipulations  and remembering that GA is a
noncommutative algebra one can construct the following linear
equation for $\psi_{+}$,
\begin{equation}\label{roteq}
R\psi_{+}\e{3}=L\psi_{+},
\end{equation}
where
\[
\begin{split}
R=&E[2\eta U\Ie{3}-k^{-2}\gamma_1\e{1}(k_x+k_y\e{12})^2\e{34}],\\
L=&(E^2+U^2-k^2)I+k^{-2}\eta U\gamma_1\e{1}(k_x+k_y\e{12})^2\e{4}.
\end{split}
\]
The Eq.~\eqref{roteq} can be rewritten in a form of  rotor
equation
\begin{equation}\label{roteqBG}
\psi_{+}\e{3}\tilde{\psi}_{+}=LR^{-1}\equiv\ba,
\end{equation}
where $\psi_{+}$ represents the rotor in 3D space and the tilde
denotes the reversion operation. The vector $\ba$ simplifies to
\begin{equation}\label{baxi}
\begin{split}
\ba&=\frac{\gamma_1\xi}{Ek^2}\,\bk\e{2}\bk+\frac{\eta U(2\xi+1)}{E}\,\e{3},\\
 \xi&=\frac{E^2-k^2-U^2}{4U^2+\gamma_1^2}.
 \end{split}
\end{equation}
\noindent The geometric product $\bk\e{2}\bk$  in polar
coordinates simplifies to $\bk\e{2}\bk/k^2=\sin(2\varphi)\,\e{1}-
\cos(2\varphi)\,\e{2}$, where $\varphi$ is the angle between
$\e{1}$ and~$\bk$. Figure~\ref{f:grapheneAxesab} shows geometric
interpretation of equation~\eqref{roteqBG}.  The initial vector
$\e{3}$, which is perpendicular to graphene plane and coincides
with quasispin quantization axis, is brought to the final vector
$\ba=\psi_{+}\e{3}\tilde{\psi}_{+}$ with the half-angle
rotor~$\psi_{+}$, where $\psi_{+}^{-1}=\tilde{\psi}_{+}$ (see
Appendix~\ref{appendOp}). One can distinguish  two extreme cases.
When the interlayer interaction is switched off, i.~e. when
$\gamma_1=0$, then $\ba\|\e{3}$ and the rotations can be performed
only around $\e{3}$ axis. In opposite case, when $U=0$, the final
vector $\ba$ lies in the graphene plane, so that the rotation
angle in this case is $\theta=\pi/2$, Fig.~\ref{f:grapheneAxesab}.
In general case, $\theta$ is determined by competition between the
interlayer interaction strength $\gamma_1$ and  external voltage
$2U$ between the layers.

For  $\psi_{+}$ to be a rotor it should be normalized,
$\psi_{+}\tilde{\psi}_{+}=1$. This guarantees that  $\psi_{+}$
brings  $\e{3}$ to a unit vector $\hat{\ba}=\ba/|\ba|$.
Mathematically the requirement that  length remains constant is
expressed by condition $\hat{\ba}^2= 1$, or in a full form
\begin{equation}\label{conda}
\hat{\ba}^2=\frac{(E^2-k^2+U^2)^2+U^2\gamma_1^2}{E^2(4U^2+\gamma_1^2)}=
1.
\end{equation}
If this fourth order  polynomial equation is solved  with respect
to energy $E$ we shall get the spectrum~\eqref{spectrum} of the
bilayer graphene.  From this we conclude that all possible vectors
$\hat{\ba}^{(i)}$ in the rotor equation
\begin{equation}
\psi_{+}^{(i)}\e{3}\tilde{\psi}_{+}^{(i)}=\hat{\ba}^{(i)}
\end{equation}
are related with the dispersion $E^{(i)}$ of the $i$-th band and
thus the vectors $\hat{\ba}^{(i)}$  represent the points on a unit
sphere.

The knowledge of $\hat{\ba}^{(i)}=\ba^{(i)}/|\ba^{(i)}|$ and
condition~\eqref{conda} allow to construct the rotor for the
$i$-th energy band,
\begin{equation}\label{psiplus}
\psi^{(i)}_{+}=\ee^{\mathcal{B}^{(i)}\theta^{(i)}/2}=
\cos\frac{\theta^{(i)}}{2}+\mathcal{B}^{(i)}\sin\frac{\theta^{(i)}}{2}.
\end{equation}
Here $\mathcal{B}^{(i)}=\e{3}\wedge\hat{\ba}^{(i)}$ is the
bivector that determines the unit oriented plane where the
rotation takes plane by angle $\theta^{(i)}$,
Fig.~\ref{f:grapheneAxesab}. The cosine of the angle
$\theta^{(i)}$ between $\e{3}$ and $\hat{\ba}^{(i)}$ can be found
from the inner GA product
$\cos\theta^{(i)}=\e{3}\cdot\hat{\ba}^{(i)}$,
\begin{equation}\label{riBG}
\begin{split}
&c^{(i)}\equiv\cos\theta^{(i)}=\\
&\eta\frac{U(2E^{(i)2}-2k^2+2U^2+\gamma_1^2)}
{\sqrt{4U^2+\gamma_1^2}\,\sqrt{(E^{(i)2}-k^2+U^2)^2+U^2\gamma_1^2}}\,.
\end{split}
\end{equation}
Elementary algebraic calculations allow to rearrange the
rotor-spinor \eqref{psiplus} to the following form
\begin{equation}\label{psipBG}
\psi_{+}^{(i)}=\frac{1}{\sqrt{2}}\Big(\sqrt{1+c^{(i)}}+(1-c^{(i)})\frac{\e{3}\bk\e2\bk}{k^2}\Big),
\end{equation}
which in the expanded form  consists of scalar and two bivectors,
$\e{23}$ and $\e{31}$, i.~e. it consists of even blades only.

The knowledge of $\psi_{+}^{(i)}$ allows to find the second
spinor-rotor $(I\psi_{-}^{(i)})$. It can be expressed through
$\psi_{+}$ if the first equation of the system~\eqref{coupled} is
used,
\begin{equation}\label{Ipsim}
(I\psi_{-}^{(i)})=\frac{\bk}{k^2}\Big(\eta
E^{(i)}\psi_{+}^{(i)}\e{3}-U\e{3}\psi_{+}^{(i)}\Big),
\end{equation}
which can be rearranged to the following form
\begin{equation}\label{psimBG}
\begin{split}
\psi_{-}^{(i)}=&\frac{\sqrt{1+c^{(i)}}}{\sqrt{2}\,k}\Big(U-E^{(i)}\eta\Big)\ee^{-\e{12}\,\varphi}\e{24}+\\
&\frac{\sqrt{1-c^{(i)}}}{\sqrt{2}\,k}\Big(U+E^{(i)}\eta\Big)\ee^{\e{12}\,\varphi}\e{34}.
\end{split}
\end{equation}

The total eigenmultivector $\psi^{(i)}$ is a sum \eqref{psipBG}
and \eqref{psimBG},
\begin{equation}\label{psipm}
\psi^{(i)}=\psi_{+}^{(i)}+\psi_{-}^{(i)}.
\end{equation}
It satisfies the eigenequation~\eqref{eigenEq}. In calculating the
physical averages the eigenmultivector $\psi^{(i)}$ must be
normalized. The square of the normalization constant is
\begin{equation}\label{normBG}
N^{(i)2}=(E^{(i)2}+k^2+U^2-2\eta Uc^{(i)} E^{(i)})/k^2.
\end{equation}

In conclusion, the eigenmultivectors have been expressed through
band energies $E^{(i)}$ and cosines $c^{(i)}=\cos\theta^{(i)}$ of
the rotation angle $\theta$ shown in Fig.~\ref{f:grapheneAxesab}.
The following sign rules should be applied to the
eigenmultivectors of individual bands:
\begin{equation}\label{bandrules}
\psi^{(i)}=
\begin{cases}
&\psi^{(1)}\quad\text{and }E^{(i)}\rightarrow -E^{(1)},\\
&\psi^{(2)}\quad\text{and }E^{(i)}\rightarrow -E^{(2)},\text{ and } c^{(i)}\rightarrow -c^{(2)},\\
&\psi^{(3)}\quad\text{and }\ c^{(i)}\rightarrow -c^{(3)},\\
&\psi^{(4)}\quad \text{Eq.}~\eqref{psipm},\\
\end{cases}
\end{equation}
where $c^{(i)}$ is given by Eq.~\eqref{riBG}. We shall remind that
the superscripts $(2)$, $(4)$ label the conduction bands and
$(1)$, $(3)$ label the valence bands. Below the
eigenmultivectors~\eqref{bandrules} will be used to analyze
pseudospin, velocity and Berry's phase properties.

\section{Pseudospin}\label{sec3}

In Hilbert basis $\{\psi_{A1},\psi_{A2},\psi_{B1},\psi_{B2}\}$ the
pseudospin  components are described by matrices
\begin{equation}
\hat{P}_x=\left[
\begin{array}{cc}
0&\hat{1}\\ \hat{1}&0
\end{array}
\right],\ \hat{P}_y=\ii\left[
\begin{array}{cc}
0&-\hat{1}\\ \hat{1}&0
\end{array}
\right],\ \hat{P}_z=\left[
\begin{array}{cc}
\hat{1}&0\\0& -\hat{1}
\end{array}
\right].
\end{equation}
In \textit{Cl}$_{3,1}$ they should be replaced by following GA
functions (see Appendix~\ref{appendRules})
\begin{equation}
\begin{split}
 P_1(\psi)&= \e{1}\psi\Ie{3},\\
 P_2(\psi)&=\e{2}\psi\Ie{3},\\
 P_3(\psi)&=\e{3}\overline{\psi}\e{3}=\e{34}\psi\e{34}.\\
\end{split}
\end{equation}
These function satisfy commutation relations of type
$P_1\big(P_2(\psi)\big)-P_2\big(P_1(\psi)\big)=2IP_3(\psi)\e{34}$
etc. From all of this follows that the average Cartesian
components of the  pseudospin  are
\begin{equation}\label{avepseudospinBG}
\begin{split}
\langle P_1\rangle&=\langle\psi^{\dagger}\e{1}\psi\Ie{3}\rangle,\\
\langle P_2\rangle&=\langle\psi^{\dagger}\e{2}\psi\Ie{3}\rangle,\\
\langle P_3\rangle&=\langle\psi^{\dagger}\e{34}\psi\e{34}\rangle,
\end{split}
\end{equation}
and the  average pseudospin and its squared module are
\begin{equation}\label{pseudospin}
\begin{split}
\mathcal{P}&=\langle P_1\rangle\e{23}+\langle
P_2\rangle\e{31}+\langle P_3\rangle\e{12},\\
|\mathcal{P}|^2&=\mathcal{P}\tilde{\mathcal{P}}=\langle
P_1\rangle^2+\langle P_2\rangle^2+\langle P_3\rangle^2.
\end{split}
\end{equation}
Note that in GA  the pseudospin is the bivector (oriented plane)
rather than  vector (oriented line), because in  GA the rotations
in an oriented plane  can be generalized to multidimensional
spaces in contrast to rotations around axis~\cite{Lounesto97}.
Formula~\eqref{pseudospin} decomposes the pseudospin into three
components (mutually perpendicular planes). Calculations give
that, in accordance with a symmetry of the problem, the pseudospin
of the $i$-th band can be expressed as
\begin{equation}\label{avespinBG}
\mathcal{P}^{(i)}=p_{\|}^{(i)}(\cos\varphi\,\e{23}+\sin\varphi\,\e{31})+p_{\bot}^{(i)}\e{12},
\end{equation}
where $\varphi$ is the angle between  $\e{1}$  and $\bk$,
Fig.~\ref{f:grapheneAxesab}.  In terms of the standard vectorial
calculus we would say that the first term in~\eqref{avespinBG}
with the amplitude $p_{\|}^{(i)}$ corresponds to dual vector
$\mathbf{P}_{\|}$ that lies in the graphene plane and is parallel
to the wave vector, $\mathbf{P}_{\|}\|\bk$, while the second term
with $p_{\bot}^{(i)}$component  represents the dual vector
$\mathbf{P}_{\bot}$ which is perpendicular to plane. In GA terms
this is expressed as
$\mathbf{P}=\mathbf{P}_{\|}+\mathbf{P}_{\bot}=\mathcal{P}I_3^{-1}$,
where $I_3=\e{1}\w\e{2}\w\e{3}$ is 3D subspace pseudoscalar.

After insertion of $\psi^{(i)}=\psi_{+}^{(i)}+\psi_{-}^{(i)}$
into~\eqref{avepseudospinBG} one finds that  for the
``Mexican-hat''-type  conduction (4) and valence (1) bands   the
amplitudes are
\begin{equation}\label{pspin14BG}
\begin{split}
p_{\|}^{(1,4)}&=\frac{2k\big(\mp\eta E^{(1,4)}-r^{(1,4)}U\big)}{E^{(1,4)2}+k^2+U^2\pm 2\eta r^{(1,4)}E^{(1,4)}U}\,,\\
p_{\bot}^{(1,4)}&=\frac{r^{(1,4)}(-E^{(1,4)2}+k^2-U^2)\mp 2\eta
E^{(1,4)}U} {E^{(1,4)2}+k^2+U^2\pm 2\eta r^{(1,4)}E^{(1,4)}U},
\end{split}
\end{equation}
where upper/lower sign is for band (1)/(4).  The amplitudes for
parabolic-like conduction (2) and valence (3) bands  are
\begin{equation}\label{pspin23BG}
\begin{split}
p_{\|}^{(2,3)}&=\frac{2k\big(\mp\eta E^{(2,3)}+r^{(2,3)}U\big)}{E^{(2,3)2}+k^2+U^2\mp 2\eta r^{(2,3)}E^{(2,3)}U}\,,\\
p_{\bot}^{(2,3)}&=\frac{r^{(2,3)}(E^{(2,3)2}-k^2+U^2)\mp 2\eta
E^{(2,3)}U} {E^{(2,3)2}+k^2+U^2\mp 2\eta r^{(2,3)}E^{(2,3)}U},
\end{split}
\end{equation}
where upper/lower sign is for band (2)/(3). Since $\eta=1$ for
$K$-valley  and $\eta=-1$ for $K^{\prime}$-valley, and the cosine
functions $r^{(i)}$  is proportional to $\eta$ [see
Eq.~\eqref{riBG}] it follows that the pseudospins of $K$ and
$K^{\prime}$ valleys are proportional to $\eta$, in other words
the pseudospins  in $K$ and $K^{\prime}$ valleys have opposite
signs (directions).

\begin{figure}[t]
\centering
\includegraphics[width=6cm]{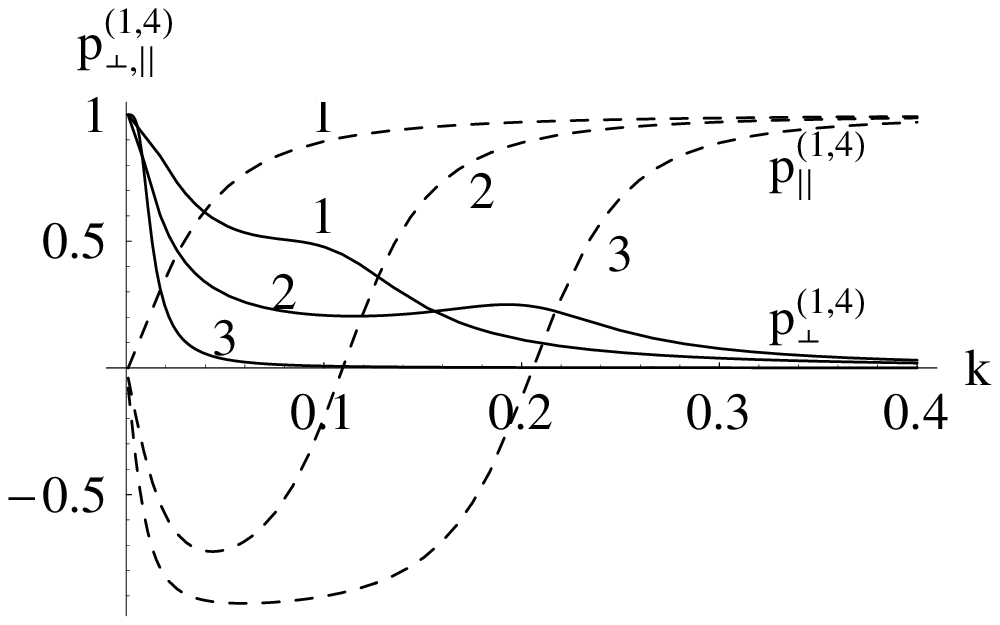}a)
\includegraphics[width=6cm]{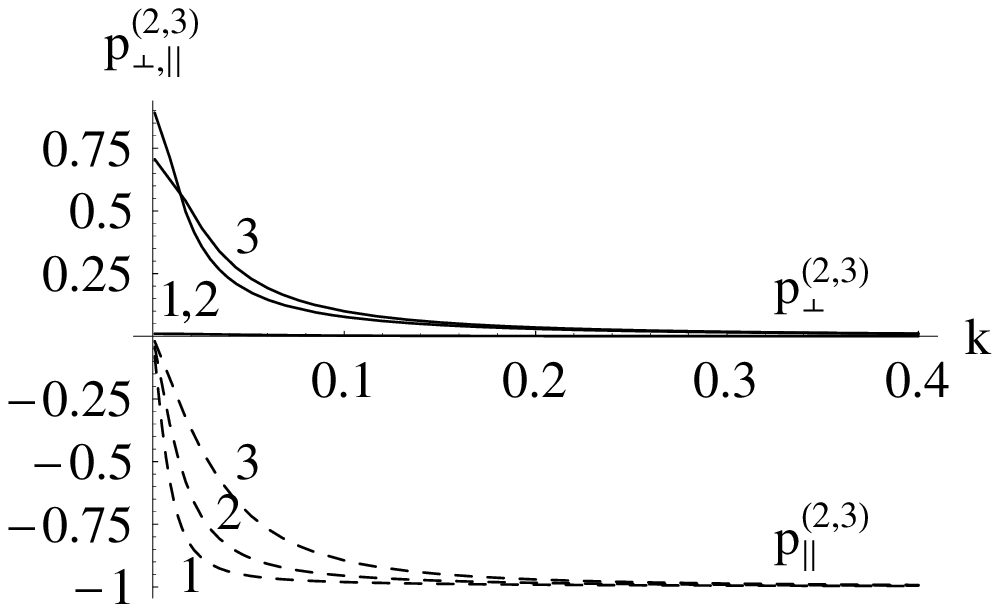}b)
\caption{Pseudospin components parallel $p_{\|}$ and perpendicular
$p_{\bot}$ to graphene surface as a function of the wave vector at
three voltages: $1-U=0.001$, $2-U=0.1$, $3-U=0.2$. a)
$p_{\|}^{(1,4)}$, and $p_{\bot}^{(1,4)}$ correspond to Mexican-hat
shaped bands  $(1)$ and $(4)$.  b) $p_{\|}^{(2,3)}$ and
$p_{\bot}^{(2,3)}$ correspond to parabolic-like bands $(2)$ and
$(3)$. In all cases $\gamma_1=0.1$.}\label{f:pspinpz}
\end{figure}

Figure~\ref{f:pspinpz} shows the amplitudes $p_{\|}^{(i)}$ and
$p_{\bot}^{(i)}$ calculated with \eqref{pspin14BG} and
\eqref{pspin23BG} as a function of wave vector~$\bk$  magnitude.
It is seen that at large wave vectors, when electron kinetic
energy predominates over interlayer coupling energy $\gamma_1$ and
voltage $U$, the pseudospin vector~$\mathbf{P}$ lies in the
graphene plane in all cases, i.~e. $\mathbf{P}_{\bot}\approx 0$.
Then $\mathbf{P}\approx\mathbf{P}_{\|}$ and $\mathbf{P}$ is either
parallel [bands (1) and (4)] or antiparallel [bands (2) and (3)]
to~$\bk$. At small $\bk$ values, or when either $\gamma_1$ or
applied voltage $2U$ begins to predominate, the component
$p_{\bot}^{(i)}$ perpendicular to graphene plane  may be large. At
$U=0$ the parallel component is
\begin{equation}
p_{\|}^{(i)}=\pm\frac{2\eta E_0^{(i)}k}{E_0^{(i)2}+k^2},
\end{equation}
where plus sign is for bands (3,4) and  minus sign is for bands
(1,2), and $E_0^{(i)}\equiv E^{(i)}(U=0)$. If $k\rightarrow 0$
then $p_{\|}^{(i)}\rightarrow 0$.

Figure~\ref{f:pspinpz}b  shows that for parabolic-like bands when
$U\rightarrow 0$ we have $p_{\bot}^{(2,3)}=0$ (note that curve 1
coincides with the horizontal axis), and the pseudospin vector
totally lies in the graphene plane and is either parallel (in
$K$-valley) or antiparallel (in $K^{\prime}$-valley) to $\bk$ for
all wave vectors (energies). Since
$E_0^{(i)}=\pm(\gamma_1/2)\pm\sqrt{k^2+(\gamma_1/2)^2}$ when
$U=0$, the point $\bk =0$ appears to be singular for bands having
``Mexican-hat'' shape. Therefore, it is convenient to introduce a
tilt angle $\vartheta$, which is defined as the angle between
pseudospin vector $\mathbf{P}$ and graphene plane. The tangent of
the tilt angle then is $\tan\vartheta=p_{\bot}/p_{\|}$.

\begin{figure}[t]
\centering
\includegraphics[width=6cm]{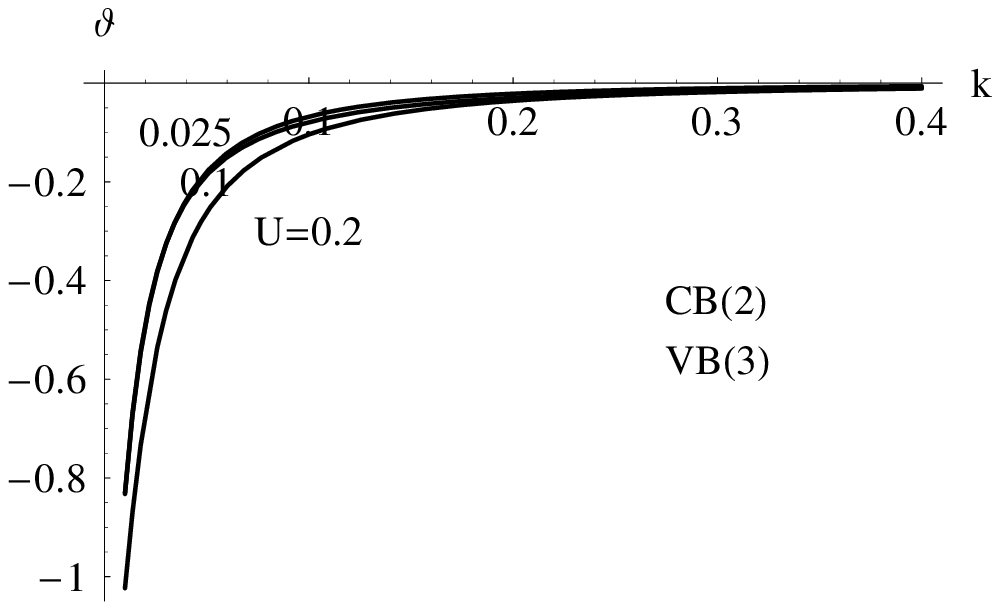}a)
\includegraphics[width=6cm]{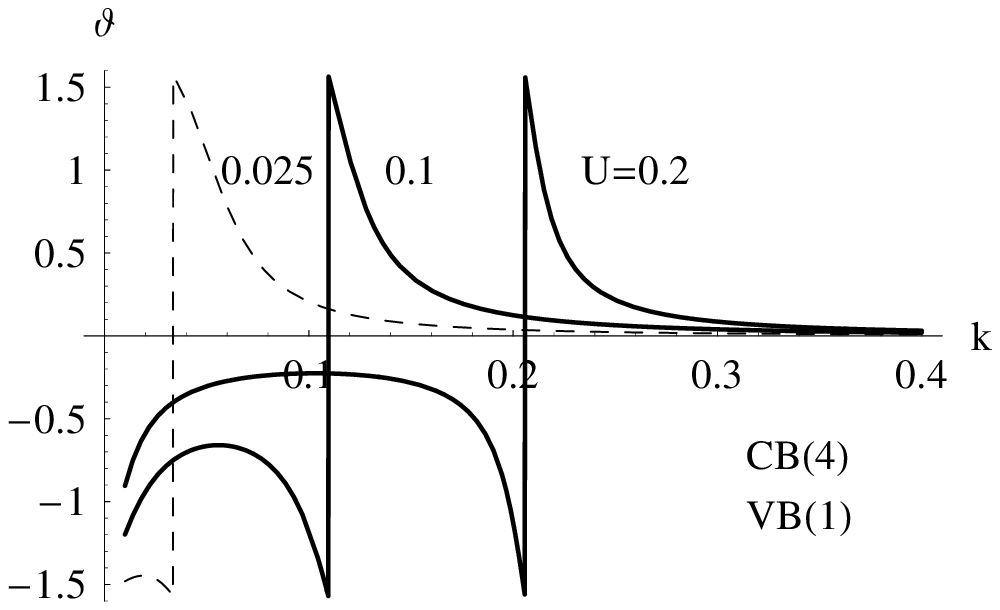}b)
\caption{The tilt angle $\vartheta$ of the pseudospin vector as a
function of the normalized wave vector $k$ for the conduction,
CB(2) and CB(4),  and valence, VB(1) and VB(3), bands at different
interlayer voltages: $2U=0.05;\ 0.2;\ 0.4$. $\gamma_1=0.1$.
}\label{f:ptiltBG}
\end{figure}

Figure~\ref{f:ptiltBG} shows the dependence of $\vartheta$ on the
wave vector magnitude when the external voltage between the layers
is $2U=0.05;0.2;0.4$. At high wave vector values the angle
$\vartheta$ goes to zero, i.~e., as mentioned, the pseudospin
vector lies in the  graphene plane. The steps in the
Fig.~\ref{f:ptiltBG}b are related with extremal points in the
``Mexican-hat'' energy dispersion, Fig.~\ref{f:spectra1234}b. At
$U=0$ and $k=0$ the step overlaps with the ordinate axis.

\section{Charge carrier velocity}\label{sec4}
The velocity in quantum mechanics is determined by vectorial
operator $\hat{\mathbf{v}}=\partial\hat{H}/\partial\bk$. The
velocity components for Hamiltonian~\eqref{hamBG} are
\begin{equation}
\hat{v}_x=\eta\left[
\begin{array}{cc}0&\hat{1}\\ \hat{1}&0\end{array}
\right],\quad
 \hat{v}_y=\ii\eta\left[
\begin{array}{cc}0&-\hat{1}\\ \hat{1}&0\end{array}
\right].
\end{equation}
After mapping to GA with the replacement rules from
Appendix~\ref{appendRules} we find the corresponding formulas
\begin{equation}\label{vxy}
\hat{v}_x|\psi\rangle\,\longrightarrow\,\eta\e{1}\psi\Ie{3},\quad
\hat{v}_y|\psi\rangle\,\longrightarrow\,\eta\e{2}\psi\Ie{3}.
\end{equation}
These components consist of even grade GA elements.  GA
expressions~\eqref{vxy} can also be obtained  directly by taking
partial derivatives of GA Hamiltonian~\eqref{HBG} with respect to
wave vector components:
\begin{equation}
\begin{split}
v_x(\psi)&=\frac{\partial H(\psi)}{\partial k_x}=\eta\e{1}\psi\Ie{3},\\
v_y(\psi)&=\frac{\partial H(\psi)}{\partial
k_y}=\eta\e{2}\psi\Ie{3}.
\end{split}
\end{equation}
The average electron velocity then is
\begin{equation}\label{avev}
\langle\mathbf{v}\rangle=\langle\psi^{\dagger}v_x(\psi)\rangle\e{1}+
\langle\psi^{\dagger}v_y(\psi)\rangle\e{2}.
\end{equation}
By convention,  in GA the angular brackets on the right-hand side
indicate that the scalar part should be taken. The velocity can
also be defined in a coordinate free way if the following vector
differential operator that acts in graphene plane is introduced
\begin{equation}\label{bnabla}
\nabla_{\bk}=\e{1}\frac{\partial}{\partial
k_x}+\e{2}\frac{\partial}{\partial k_y}.
\end{equation}
Then the average electron velocity can be written in the following
coordinate-free form
\begin{equation}\label{avevel}
\langle\mathbf{v}\rangle=\dot{\nabla}_{\bk}\langle\psi^{\dagger}\dot{H}(\psi)\rangle,
\end{equation}
where the dagger operation is defined by formula~\eqref{dagger} in
the Appendix~\ref{appendOp}. Since $\e{i}\psi\ne\psi\e{i}$, the
operator $\nabla_{\bk}$ cannot be pushed through spinor
$\psi^{\dagger}$. The pair of the overdots in~\eqref{avevel}
indicates that only the partial derivatives $\partial/\partial
k_{x}$ and $\partial/\partial k_{y}$ act on the Hamiltonian
directly.

\begin{figure}
\centering
\includegraphics[width=7cm]{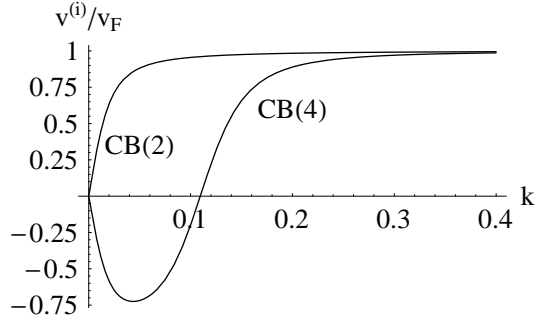}
\caption{Average electron velocity vs wave vector in conduction
bands of BLG. $v_F$ is the asymptotic velocity when
$k\rightarrow\infty$. $U=0.1$.}\label{f:velBG}
\end{figure}

Calculations  with~\eqref{avev} or \eqref{avevel} give the
following   general expression for the $i$-th band average
velocity
\begin{equation}\label{avel}
\begin{split}
&\langle\mathbf{v}^{(i)}\rangle=v^{(i)}\big(\e{1}\cos\varphi+\e{2}\sin\varphi\big),\\
&v^{(i)}=\pm \frac{2k(\pm E^{(i)}+\eta
r^{(i)}U)}{E^{(i)2}+k^2+U^2\pm
2\eta E^{(i)}r^{(i)}U},\\
\end{split}
\end{equation}
where the sign in the amplitude $v^{(i)}$ depends on band index.
It is seen that $\langle\mathbf{v}^{(i)}\rangle\|\bk$, in
accordance with the isotropic character of energy bands. If the
velocity~\eqref{avel} is expanded in a full form the dependence on
valley index $\eta$ vanishes. This means that in both $K$ and
$K^{\prime}$ valleys the electron velocity magnitude and direction
are the same. Figure~\ref{f:velBG} illustrates the average
velocity of electron in the conduction band calculated
with~\eqref{avel}, where the change of sign in $v^{(4)}$ is
related with the ``Mexican-hat'' character of the conduction band
dispersion. The same velocities are found for valence bands.

It can be shown that the average velocity can be obtained in a
simpler way, directly from the respective band,
\begin{equation}
\langle\mathbf{v}^{(i)}\rangle=\nabla_{\bk}E^{(i)}= \frac{\partial
E^{(i)}}{\partial k_x}\e{1}+\frac{\partial E^{(i)}}{\partial
k_y}\e{2},
\end{equation}
where $\nabla_{\bk}$ is the vectorial derivative~\eqref{bnabla}.

\section{Berry phase}\label{sec5}
At  low free carrier  energies the Berry phase of BLG is equal to
$2\pi$~\cite{McCann06}. As we shall see, in the presence of the
interlayer potential the Berry phase no longer remains a multiple
of $\pi$.

The Berry phase $\Gamma$ can be defined through a sum of matrix
elements between adjacent points~\cite{Park11}
\begin{equation}\label{Berry}
\Gamma=-\ii\lim_{N\rightarrow\infty}\sum_{j=0}^{N-1}\log\langle
\psi_j|\psi_{j+1}\rangle,
\end{equation}
where $j$ is the state index. The sum in~\eqref{Berry} is closed
so that the  initial and final states coincide,
$|\psi_0\rangle=|\psi_{N}\rangle$. If the replacement rule  for
the matrix element $\langle\varphi|\psi\rangle$  (see
Appendix~\ref{appendRules}) is applied to Eq.~\eqref{Berry} then
in \textit{Cl}$_{3,1}$ algebra the Berry phase will read
\begin{equation}\label{BerryGA}
\begin{split}
&\Gamma=-\e{12}\lim_{N\rightarrow\infty}\sum_{j=0}^{N-1}\log\langle j,j+1\rangle,\\
&\langle j,j+1\rangle=\langle\psi_j^{\dagger}\psi_{j+1}\rangle-
\langle\psi_j^{\dagger}\psi_{j+1}\,\e{12}\rangle\e{12}.
\end{split}
\end{equation}
The main contribution to the Berry phase  comes from the last term
with two bivectors $\e{12}$. For simplicity we shall assume that
the length of the wave vector~$\mathbf{k}$ is constant so that the
loop around the center of $K$-valley becomes the circle, and
therefore only the angle $\varphi$ varies. The spinor state $\psi$
in $K$-valley is given by equations
\eqref{psipm}-\eqref{bandrules}, where the angle
$\varphi=\text{arctan}(k_y/k_x)$ in~\eqref{BerryGA} should be
replaced by either $\varphi_j$ or $\varphi_j+\delta\varphi_j$. We
shall assume that the increment $\delta\varphi_j$ is small enough
so that trigonometric functions in spinors  can be expanded in
series. Then to the first order in $\delta\varphi_j$ we find that
$\langle\psi_j^{\dagger}\psi_{j+1}\rangle\approx 1$ and
$\langle\psi_j^{\dagger}\psi_{j+1}\,\e{12}\rangle\approx-\beta^{(i)}\delta\varphi_j$,
so that the matrix element  between  the state $j$ (angle
$\varphi_{j}$) and  neighboring state $j+1$ (angle
$\varphi_{j}+\delta\varphi_j$)  can be approximated by
\begin{equation}
\langle
j,j+1\rangle^{(i)}\approx1+\beta^{(i)}\delta\varphi_j\,\e{12},
\end{equation}
where the scalar coefficient is
\begin{equation}
\beta^{(i)}\approx 1-
\frac{k^2c^{(i)}}{E^{(i)2}+k^2+U^2-2E^{(i)}Uc^{(i)}}.
\end{equation}
It can be shown that $\beta^{(1)}=\beta^{(4)}$ and
$\beta^{(2)}=\beta^{(3)}$, i.~e., we have the same Berry phase for
conduction as well  as for valence band of similar shape. Since at
small increments of angle we have approximately $\log\langle
j,j+1\rangle^{(i)}\approx\beta^{(i)}\delta\varphi_j\,\e{12}$, from
the above expressions it follows that in the limit
$N\rightarrow\infty$ the Berry phase becomes
\begin{equation}\label{Gamma2}
\Gamma^{(i)}=\lim_{N\rightarrow\infty}\Big(-\e{12}\beta^{(i)}
\sum_{j=0}^{N-1}\delta\varphi_j\,\e{12}\Big)=2\pi\beta^{(i)}.
\end{equation}
Since $c^{(i)}$ is proportional to $U$ (see Eq.~\eqref{riBG}), we
find that $\beta^{(i)}\rightarrow 1$ when $U\rightarrow 0$, and as
a result the expression~\eqref{Gamma2} reduces to the value
$\Gamma^{(i)}=2\pi$, which is independent of the band index $i$,
in agreement with the experiment \cite{Novoselov06}. However, when
$U\ne 0$ then $\beta^{(i)}\ne 1$ and a simple $n\pi$-type nature
of the Berry phase vanishes.

\begin{figure}
\centering
\includegraphics[width=6.3cm]{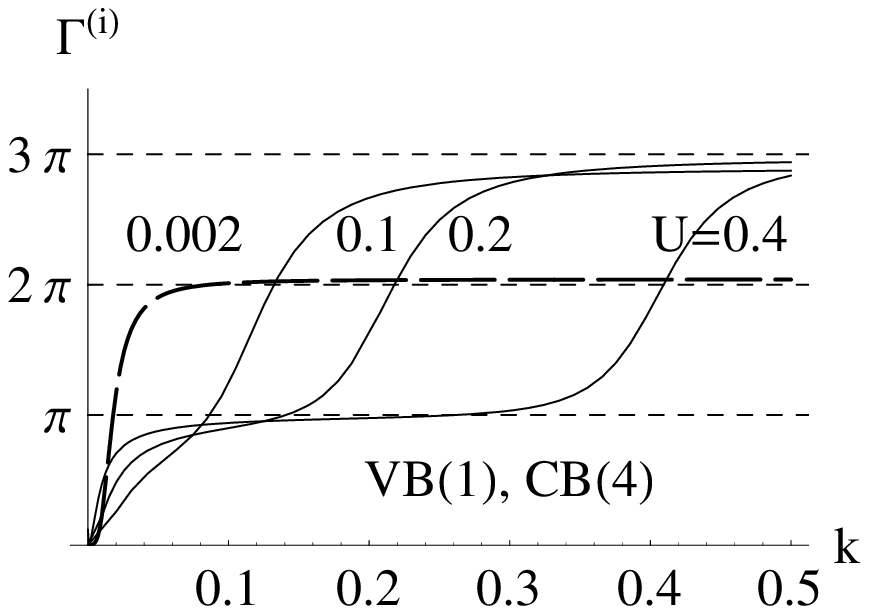}a)
\includegraphics[width=6.5cm]{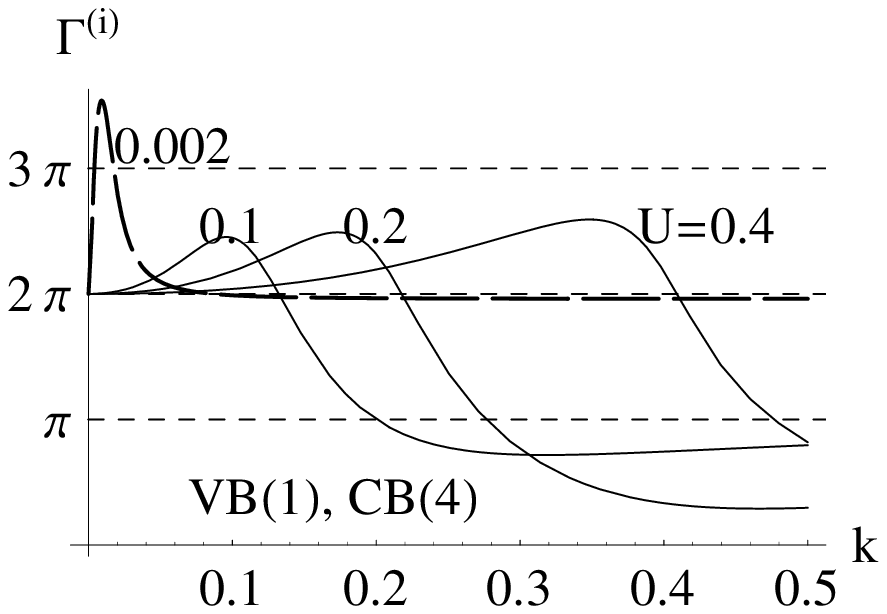}b)
\caption{Berry phase for bilayer graphene in gap forming
valence~$i=1$ and conduction~$i=4$ bands.  a) $\eta=+1$ and b)
$\eta=-1$. $U=0.002;\ 0.1;\ 0.2;\ 0.4$.
$\gamma_1=0.1$.}\label{f:BerryBG14}
\end{figure}

Figures~\ref{f:BerryBG14} show the dependence of the Berry phase
in $K$ ($\eta=1$) and $K^{\prime}$ ($\eta=-1$) valleys. When $k=0$
the minimum/maximum of the energy is located in the center of the
valley and the Berry phase in all cases is equal to either $2\pi$
or $0$, i.~e. it is a multiple of $2\pi$, in agreement with the
arguments of Ref.~\cite{Park11}. When $k\ne 0$ the variation of
the Berry phase vs $k$ is different for different valleys and
bands. In the gap-forming energy bands $(1,4)$ the shift of energy
maxima/minima in the presence of $U$ is directly reflected in the
Berry phase as a fast switching from $n\pi$ to $(n\pm 1)\pi$ value
as featured by Fig.~\ref{f:BerryBG14}a. In case of remote,
parabolic-like bands $(2,3)$ the interlayer potential $U$ does not
change the location of critical points in the Brillouin zone. By
this reason the phases $\Gamma^{(2)}$ and $\Gamma^{(3)}$ for bands
$(2,3)$ were found to change slowly with $k$ magnitude.

\section{Discussion}
Analytical expressions for pseudospin, velocity and Berry phase
were obtained for both $K$ and $K^{\prime}$ valleys of the bilayer
graphene in the presence of external voltage applied between the
two layers. The obtained formulas are expressed in terms of band
eigenenergies.  It is shown how the eigenequation for bilayer
graphene can be reduced to rotor equation in 3D Euclidean space in
terms of eigenenergies and the rotation angle that connects
pseudospin quantization axis $\e{3}$ with the final axis
determined by the Hamiltonian of the problem. This property allows
the solutions to be interpreted as points on a unit sphere in 3D
space, where the North pole represents the quantization axis. The
eigenmultivectors of bilayer Hamiltonian  were found to be the
rotors which connect the North pole with other points (solutions)
on the sphere. In particular the paper shows how one can construct
the needed rotors with \textit{Cl}$_{3,1}$ algebra. Since GA
rotors  are connected with spinors they can be applied  to
calculate various physical properties of graphene. Since GA
approach is coordinate-free the formulas found in this way appear
to be rather compact and may be interpreted geometrically. In
particular, it is shown that the external voltage destroys $n\pi$
character of the Berry phase, where $n$ is the integer.

\textit{Cl}$_{3,1}$  algebra  has the signature $(+,+,+,-)$ and is
applied to relativity theory, including the Dirac
equation~\cite{Doran03}. Here, \textit{Cl}$_{3,1}$ algebra was
used to  analyze a nonrelativistic stationary Schr{\"o}dinger
equation, where time is a parameter rather then the fourth
coordinate $\e{4}$ in 4D spacetime of relativity. In the
considered problem the basis vector $\e{4}$ plays an auxiliary
role that allows to divide the spinor into even and odd parts with
respect to spatial inversion. As shown in~\cite{Dargys13a}, to
describe the monolayer graphene the smallest algebra is
\textit{Cl}$_{3,0}$ which represents 3D Euclidean space and  has
$2^3=8$ basic elements in the Clifford space. The double layer
graphene requires two times larger \textit{Cl}$_{3,1}$ algebra.
Its irreducible representation is made up of complex $4\times 4$
matrices as follows from the $8$-periodicity
table~\cite{Lounesto97}. There are more algebras that can be
represented by such matrices, for example, \textit{Cl}$_{1,3}$ and
\textit{Cl}$_{4,1}$. Which of the algebras is best suited for
description of 2D materials in GA terms at this moment is not
clear and more investigations are needed. If, in addition, one
wants to take into account the intrinsic electron spin then one
must address to $8\times 8$ matrix representation and,
respectively, larger Clifford algebra. Finally, the same results,
in principle, can be obtained within the standard Hilbert space
approach. However, the geometric algebra approach is superior
since it is coordinate-free, takes the quantization axis into
account explicitly, is endowed with geometric interpretation and,
in general, is universal, i.e. the same mathematical machinery can
be applied  from the simplest Newton mechanics to quantum
relativity and cosmology~\cite{Doran03}.

\section{Appendix: \textit{Cl}$_{3,1}$ algebra \label{appendix}}
\subsection{Matrix representation\label{appendMat}}
The squares of basis vectors $\e{i}$  of \textit{Cl}$_{3,1}$
algebra satisfy $\e{1}^2=\e{2}^2=\e{3}^2=1$ and $\e{4}^2=-1$. The
following $4\times 4$ complex matrix representation of $\e{i}$
will be used in decomposing the Hamiltonian, pseudospin and
velocity matrices
\begin{equation}\label{e1e2}
\hat{e}_{1}=\left[
\begin{array}{cc}
0&\hat{1}\\
\hat{1}&0\\
\end{array}
\right], \quad
 \hat{e}_{2}=\ii\left[
\begin{array}{cccc}
0&-\hat{1}\\
\hat{1}&0\\
\end{array}
\right],
\end{equation}
\begin{equation}\label{e3e4}
\hat{e}_{3}=\left[
\begin{array}{cc}
\hat{\sigma}_y&0\\
0&-\hat{\sigma}_y\\
\end{array}
\right], \quad
 \hat{e}_{4}=\ii\left[
\begin{array}{cccc}
\hat{\sigma}_z&0\\
0&-\hat{\sigma}_z\\
\end{array}
\right],
\end{equation}
where $\ii=\sqrt{-1}\,$. $\hat{1}$ is $2\times 2$ unit matrix and
$\hat{\sigma}_x$, $\hat{\sigma}_y$, $\hat{\sigma}_z$  are Pauli
matrices.  The products of matrices $\hat{e}_i$ is equivalent to
geometric products of basis vectors $\e{i}$ in GA. For example,
the product of all four matrices gives matrix representation of GA
pseudoscalar $I\equiv\e{1234}$,
\begin{equation}
\hat{I}=\hat{e}_1\hat{e}_2\hat{e}_3\hat{e}_4=\ii\left[
\begin{array}{cc}
-\hat{\sigma}_x&0\\
0&\hat{\sigma}_x\\
\end{array}
\right].
\end{equation}
The other matrices that appear in the Hamiltonian~\eqref{hatH} are
\begin{equation}\label{e3e1}
\hat{e}_{31}\equiv\hat{e}_{3}\hat{e}_{1}=\left[
\begin{array}{cc}
0&\hat{\sigma}_y\\ -\hat{\sigma}_y&0
\end{array}\right],
\end{equation}
\begin{equation}\label{Ie2}
\hat{I}\hat{e}_2=-\left[
\begin{array}{cc}
0&\hat{\sigma}_x\\ \hat{\sigma}_x&0
\end{array}\right],\quad
\hat{I}\hat{e}_3=\left[
\begin{array}{cc}
\hat{\sigma}_z&0\\ 0&\hat{\sigma}_z
\end{array}\right].
\end{equation}
The square of $\hat{I}$  gives the $4\times 4$ unit matrix with
negative sign. This is equivalent to $I^2=-1$ in GA. Similarly,
the pairwise products of matrices give the images  of bivectors,
for example, the bivector $\e{31}$ is represented by
matrix~\eqref{e3e1}. The products of three different matrices
generate the  trivectors. For example, trivectors $\e{134}=\Ie{2}$
and $-\e{124}=\Ie{3}$ are represented by matrices~\eqref{Ie2}.
All in all, there are $2^4=16$ different matrices that represent
all basis elements of \textit{Cl}$_{3,1}$ algebra.

\subsection{Replacement rules\label{appendRules}}
The following mapping is assumed between the complex spinor
$|\psi\rangle$ in a form of column and GA spinor $\psi$,
\begin{equation}\label{psipsi}
\begin{split}
 |\psi\rangle&=\left[
\begin{array}{c}
\ \,a_0+\ii a_3\\
- b_3+\ii b_0\\
-b_2-\ii b_1\\
-a_1+\ii a_2
\end{array}\right]
\longleftrightarrow\,\psi=a_0+a_1\e{23}-\\
&a_2\e{31}+a_3\e{12}-b_0I-b_1\e{14}+b_2\e{24}+b_3\e{34},
\end{split}
\end{equation}
where $a_i$'s and $b_i$'s are scalars.

The knowledge of matrix representation of basis elements
[equations \eqref{e1e2} and \eqref{e3e4}] and spinor mapping
rule~\eqref{psipsi} along with the idempotents of
\textit{Cl}$_{3,1}$ algebra~\cite{Ablamowicz03} allow to construct
the following main replacement rules between the action of
matrices on column vectors  and \textit{Cl}$_{3,1}$ multivector
functions,
\begin{equation}\label{repRulesAA}
\begin{split}
\hat{e}_i|\psi\rangle&\longleftrightarrow\e{i}\psi\Ie{3},\\
\hat{e}_{ij}|\psi\rangle&\longleftrightarrow\e{ij}\psi,\\
\hat{I}|\psi\rangle&\longleftrightarrow I\psi,\\
\hat{I}\hat{e}_i|\psi\rangle&\longleftrightarrow\Ie{i}\psi\Ie{3},
\end{split}
\end{equation}
where $i,j=1,2,3,4$. Also, additional replacement rules may be
useful
\begin{equation}\label{repRulesA}
\begin{split}
\ii|\psi\rangle&\longleftrightarrow I\psi\e{34},\\
\langle\psi|\psi\rangle&\longleftrightarrow\langle\psi^{\dagger}\psi\rangle,\\
\langle\varphi|\psi\rangle&\longleftrightarrow\langle\varphi^{\dagger}\psi\rangle-\langle\varphi^{\dagger}\psi\,\e{12}\rangle\e{12}.
\end{split}
\end{equation}
The angled brackets in GA, for example $\langle M\rangle$,
indicate that only the scalar part of the multivector~$M$ should
be taken. The meaning of the dagger operation is defined below.

\subsection{Spatial inversion, reversion and dagger operations\label{appendOp}}

The spatial inversion  (denoted by overbar) changes signs of all
spatial vectors to opposite but  leaves ``time'' vector $\e{4}$
invariant: $\inv{\e{}}_i=-\e{i}$ if $i$=1,2,3 and
$\inv{\e{}}_4=\e{4}$. In \textit{Cl}$_{3,1}$,  inversion of a
general multivector $M$ is defined by
\begin{equation}
\inv{M}=-\e{4}M\e{4}.
\end{equation}
Properties of inversion:  $\inv{M_1+M_2}=\inv{M}_1+\inv{M}_2$, and
$\inv{M_1M_2}=\inv{M}_1\,\,\inv{M}_2$. The spatial inversion
allows to divide the general spinor~\eqref{psipsi} into even and
odd parts, $\psi=\psi_{+}+\psi_{-}$, where
\begin{equation}
\begin{split}
\psi_{+}&=(\psi+\inv{\psi})/2=a_0+a_1\e{23}-a_2\e{31}+a_3\e{12},\\
\psi_{-}&=(\psi-\inv{\psi})/2=-b_0I-b_1\e{14}+b_2\e{24}+b_3\e{34},
\end{split}
\end{equation}
which satisfy $\inv{\psi}_{+}=\psi_{+}$ and
$\inv{\psi}_{-}=-\psi_{-}$.

The operation of reversion (denoted by tilde) reverses the order
of basis vectors in the multivector.  For example, after reversion
the bivector changes its sign
$\widetilde{\mathbf{e}}_{12}=\e{21}=-\e{12}$. The reversion
usually is used to find the norm of the blade. If applied to
$\psi$ the reversion  gives the difference between coefficients of
even and odd parts of the spinor $\psi$,
\begin{equation}
\begin{split}
\langle\tilde{\psi}\psi\rangle&=\tilde{\psi}_{+}\psi_{+}+\tilde{\psi}_{-}\psi_{-}=\\
&(a_0^2+a_1^2+a_2^2+a_3^2)-(b_0^2+b_1^2+b_2^2+b_3^2).
\end{split}
\end{equation}

The dagger operation is a combination of the reversion and
inversion
\begin{equation}\label{dagger}
\psi^{\dagger}=-\e{4}\tilde{\psi}\e{4}.
\end{equation}
If applied to a general bispinor it allows to find the square of
the module
\begin{equation}
\begin{split}
\langle\psi^{\dagger}\psi\rangle&=\psi_{+}^{\dagger}{\psi}_{+}+\psi_{-}^{\dagger}{\psi}_{-}=\\
&(a_0^2+a_1^2+a_2^2+a_3^2)+(b_0^2+b_1^2+b_2^2+b_3^2).
\end{split}
\end{equation}
Both  operations are  symmetric, e.g.,
$\langle\psi\psi^{\dagger}\rangle=\langle\psi^{\dagger}\psi\rangle$.


\begin{thebibliography}{[1]}


\bibitem{McCann06}
E.~Mc{C}ann and V.~I. Fal'ko,  Phys. Rev. Lett. {\bf 96}, 086805
(2006).

\bibitem{Kuzmenko09}
A.~B. Kuzmenko, I.~Crassee, D.~van~der {M}arel, P.~Blake, and
K.~S. Novoselov, Phys. Rev. B {\bf 80},   165406 (2009).

\bibitem{McCann13}
E.~Mc{C}ann and M.~Koshino,  arXiv
  cond-mat.mes-hall 1205.6953v2, 31 (2013).

\bibitem{Abergel10}
D.~S.~L. Abergel, V.~Apalkov, J.~Barashevich, K.~Ziegler, and
T.~Chakraborty, Adv. Phys. {\bf 59}, 261  (2010) .

\bibitem{Katsnelson12}
M.~I. Katsnelson, Graphene: {C}arbon in Two Dimensions (Cambridge
University  Press, Cambridge, 2012).

\bibitem{Lounesto97}
P.~Lounesto, Clifford Algebra and Spinors, (Cambridge University
Press,  Cambridge, 1997).

\bibitem{Doran03}
C.~Doran and A.~Lasenby, Geometric Algebra for Physicists,
(Cambridge University  Press, Cambridge, 2003).

\bibitem{Dorst11}
L.~Dorst and J.~Lasenby, (eds), Guide to Geometric Algebra in
Practice (Springer, London, 2011).

\bibitem{Hestenes92}
D.~Hestenes,  Advances in Applied Clifford
  Algebras {\bf 2}, 215  (1992).

\bibitem{Dargys13}
A.~Dargys, Physica E {\bf 47}, 47  (2013).

\bibitem{Bohmer12a}
C.~G B{\" o}hmer and L.~Corpe,  J.~Phys.A: Math. Theor. {\bf 45}
 205206 (2012).

\bibitem{Dargys13a}
A.~Dargys,  Acta. Phys. Polonica A, {\bf 124},  732 (2013).

\bibitem{Dargys14}
A.~Dargys,  Lithuanian J. Phys. {\bf 54}, 33 (2014).

\bibitem{Dargys14b}
A.~Dargys and A.~Acus, 10th Int. Conf. on Clifford Algebras and
their Applications in Math. Phys., Tartu (2014).


\bibitem{Park11}
C.~H. Park and N.~Marzari,  Phys. Rev. B {\bf 84},  205440 (2011).

\bibitem{Novoselov06}
K.~S. Novoselov, E.~Mc{C}ann, S.~V. Morozov, V.~I. Fal'ko, M.~I.
Katsnelson,  U.~Zeitler, D.~Jiang,  F.~Schedin, and A.~K. Geim,
Nature Physics {\bf 21}  177  (2006).

\bibitem{Ablamowicz03}
R.~Ab{\l}amowicz, B.~Fauser, K.~Podlaski, and J.~Rembieli{\'n}ski,
Czechoslovak J. Phys. {\bf 53},  949 (2003);
arXiv:math-ph/0312015v1.

\end{thebibliography}
\end{document}